\documentclass[a4paper,10pt]{article}

\usepackage{graphics,epsfig}
\usepackage{amsmath}
\usepackage{amsfonts}
\usepackage{amssymb}

\newcommand{\be}{\begin{equation}}
\newcommand{\ee}{\end{equation}}
\newcommand{\ba}{\begin{eqnarray}}
\newcommand{\ea}{\end{eqnarray}}

\begin{document}

\title{Chemical potentials and parity breaking: the Nambu-Jona-Lasinio model}
\author{A. A. Andrianov$^{1,2}$, D. Espriu$^{2}$ and X. Planells$^{2}$\\
\small{$^1$ V. A. Fock Department of Theoretical Physics,  Saint-Petersburg State University,}\\
{\small 198504 St. Petersburg, Russia}\\
{\small $^2$ Departament d'Estructura i Constituents de la Mat\`eria and }\\
{\small Institut de Ci\`encies del Cosmos (ICCUB),
Universitat de Barcelona,}\\
{\small  Mart\'\i \ i Franqu\`es 1, 08028 Barcelona, Spain}}

\date{December 2013}

\maketitle
\begin{abstract}
We consider the ``two flavour'' Nambu--Jona-Lasinio model in the presence of a vector and an axial external 
chemical potentials and study the phase structure of the model at zero temperature. The Nambu--Jona-Lasinio 
model is often used as a toy replica of QCD and it is therefore interesting to explore 
the consequences of adding external vector and
axial chemical potentials in this model, mostly motivated by claims
that such external drivers could trigger a phase where parity could 
be broken in QCD. We are also motivated by some lattice analysis that attempt to
understand the nature of the so-called Aoki phase using this simplified model.
Analogies and differences with the expected behaviour in QCD are discussed and
the limitations of the model are pointed out. 
\end{abstract}

\vspace{-13cm}
\begin{flushright} ICCUB-13-229
\end{flushright}  
\vspace{12.25cm}

\section{Motivation}
In the last years, the possibility that parity breaks in QCD at high temperatures and/or densities has 
received a lot of attention \cite{kharzeev,anesp}. Although parity is well known to be a symmetry of strong interactions, 
there are reasons to believe that it may be broken in extreme conditions. On the one hand, theoretical 
work using effective meson lagrangians satisfying the QCD symmetries at low energies 
suggest that for some values of the vector chemical potential $\mu$ a new phase with an 
isotriplet pseudoscalar 
condensate may arise \cite{anesp}. On the other hand, thermal fluctuations in a finite volume may lead to 
large topological fluctuations that induce a non-trivial axial quark charge that could be described
in a quasi-equilibrium situation by an axial chemical potential $\mu_5$ \cite{kharzeev,aep,polikarpov}.

Checking these claims in QCD is unfortunately very difficult. For one thing, finite density 
numerical simulations in the lattice present serious difficulties \cite{latt}. A 
vector chemical potential in gauge theories like QCD cannot be easily treated and therefore 
simpler models hopefully reproducing the main features of the theory may be
useful. Needless to say, non-equilibrium effects are also notoriously difficult to study
non-perturbatively. However, an axial chemical potential is tractable on the lattice \cite{yama}
and with other methods \cite{ChernLuo}.

In the present paper we shall consider the Nambu--Jona-Lasinio model (NJL) \cite{NJL,NNJJLL,NNJJLL2,Ebert}, which
shares interesting features with QCD such as the appearance of chiral symmetry breaking. In the
NJL modelization, QCD gluon interactions among fermions are assumed to be replaced by some effective 
four-fermion couplings. Confinement is absent in the NJL model, but global symmetries can be
arranged to be identical
in both theories.

However, NJL is definitely not QCD and the present work does not attempt to draw definite conclusions on the latter theory; just to point out possible phases requiring further analysis.

Previously some authors have studied the effect of a vector chemical potential $\mu$ with three flavours \cite{ikkk} 
in the NJL model, but the consequences of including both a vector and an axial chemical potentials have not been considered so far
to our knowledge. In this work, we will incorporate both chemical potentials 
with the purpose of unravelling the landscape of different stable phases of the theory. It turns out that 
the inclusion of $\mu_5$ changes radically the phase structure of the model and shows that $\mu$ 
is not a key player in ushering a thermodynamically stable phase where parity is violated in the
NJL model, but $\mu_5$ is.

This paper is organized as follows. In Sec. 2 the NJL Lagrangian with the incorporation of $\mu$ and $\mu_5$ 
will be introduced. We describe how an effective potential is extracted when one 
introduces some effective light meson states and integrates out the fermion degrees of freedom. 
In Sec. 3 we show the gap equations of the model and the  conditions for
their stability. After that, the different stable phases of this model are presented and discussed. 
We show in Sec. 4 that a phase with an isospin singlet pseudoscalar condensate in addition to a scalar 
condensate is possible. 
It turns out that the conditions for this phase to be stable and exhibit chiral symmetry
breaking too are such that one gets an inverted mass spectrum with $m_\pi>m_{\eta_q}$ and $m_\sigma>m_{a_0}$, which 
is quite different from QCD. In  
Sec. 4 we also present the main results of this work with plots of the evolution of the scalar and 
pseudoscalar condensates together with the main features of the phase transition. 
Finally, Sec. 5 is devoted to summarize our conclusions.

\section{NJL Lagrangian with $\mu$ and $\mu_5$}
The starting point of this work is the NJL Lagrangian where we incorporate a vector and an axial chemical 
potentials $\mu$ and $\mu_5$, respectively. For two flavours and $N$ colours, we have 
\begin{equation}\label{lagNJL}
\mathcal L=\bar\psi(\partial\!\!\!\!\!\!\not\;\; + m 
- \mu\gamma_0 - \mu_5\gamma_0\gamma_5)\psi - 
\frac{G_1}N[(\bar\psi\psi)^2+(\bar\psi i\gamma_5\vec\tau\psi)^2]
-\frac{G_2}N[(\bar\psi\vec\tau\psi)^2+(\bar\psi i\gamma_5\psi)^2],
\end{equation}
with a full $U(2)_L\times U(2)_R$ chiral invariance in the case that $G_1=G_2$, while if these constants 
differ, the $U(1)_A$ symmetry breaks and only $SU(2)_L\times SU(2)_R\times U(1)_V$ remains. One may introduce 
two doublets of bosonic degrees of freedom $\{\sigma,\vec\pi\}$ 
and $\{\eta,\vec a\}$ by adding the following chiral invariant term
\begin{equation}
\Delta\mathcal L=\frac{Ng_1^2}{4G_1}(\sigma^2+\vec \pi^2)+\frac{Ng_2^2}{4G_2}(\eta^2+\vec a^2).
\end{equation}
These would be identified with their namesake QCD states (actually $\eta_q$ and $\vec a_0$ for the last two). Euclidean
conventions will be used throughout. We bosonize the model following the same procedure as in \cite{NNJJLL}.

After shifting each bosonic field with the quark bilinear operator that carries the corresponding quantum numbers, 
the Lagrangian \eqref{lagNJL} may be rewritten as
\begin{equation}
\mathcal L=\bar\psi[\partial\!\!\!\!\!\!\not\;\; + m -\mu\gamma_0 -\mu_5\gamma_0\gamma_5
+g_1(\sigma+i\gamma_5\vec\tau\vec\pi)+g_2(i\gamma_5\eta+\vec\tau\vec a)]\psi
+\frac{Ng_1^2}{4G_1}(\sigma^2+\vec \pi^2)+\frac{Ng_2^2}{4G_2}(\eta^2+\vec a^2),
\end{equation}
which shows a redundancy related to the coupling constants $g_{1,2}$ that appear attached to each doublet 
and it is eventually related to their wave function normalization. Without further ado we will take $g_1=g_2=1$.

Integration of the fermions will produce a bosonic effective potential (or free energy) and will
allow to study the different phases of the model. We will work in the mean field approximation and accordingly neglect
fluctuations. The results will be exact in the large $N$ limit.
\begin{equation}\label{V}
V_{\text{eff}}=\frac{N}{4G_1}(\sigma^2+\vec \pi^2)+\frac{N}{4G_2}(\eta^2+\vec a^2)- \text{Tr}\log\mathcal M(\mu,\mu_5),
\end{equation}
where the trace is understood to be performed in the isospin and Dirac spaces in addition to a 4-momentum integration 
of the operator in the momentum space. Throughout this article we will assume that $\mu>0$, namely we consider a baryon (as opposed
to antibaryon) finite density. The invariance under $CP$ of the action ensures that the free energy (\ref{V}) only 
depends on the modulus of $\mu$.

We also define the fermion operator
\begin{equation}\label{fermoper}
\mathcal M(\mu,\mu_5)=\partial\!\!\!\!\!\!\not\;\; + (M+\vec\tau\vec a)- \mu\gamma_0 - \mu_5\gamma_0\gamma_5
+i\gamma_5(\vec\tau\vec\pi+\eta),
\end{equation}
with the introduction of a constituent quark mass $M\equiv m+\sigma$.

In appendix \ref{trick} we show that the dependence on both vector and axial chemical potentials does not change the reality 
of the fermion determinant. However, its sign remains undetermined, and in order to ensure a positive determinant, we shall 
consider an even number of "colours"\footnote{The choice of an even number of colours, unlike QCD, is simply a technical restriction to ensure the fermion determinant to be positive definite.} $N$ so that one can safely assume
\begin{equation}\label{det+}
\det[\mathcal M(\mu,\mu_5)]=\sqrt{\det[\mathcal M(\mu,\mu_5)]^2}
\end{equation}
and hence, use the calculations in Appendix \ref{trick}. If we just retain the neutral components of the triplets, this determinant can be written in the following way
\begin{align}\label{Trlog}
\nonumber \log&\det\mathcal M(\mu,\mu_5)=\text{Tr}\log\mathcal M(\mu,\mu_5)\\
&\!\!\!\!\!\!\!\!\!=\frac18\text{Tr}\sum_{\pm}\Bigg\{\log\left [-(ik_0+ \mu)^2+(|\vec k|\pm \mu_5)^2+M_+^2\right ]
+\log\left [-(ik_0 + \mu)^2+(|\vec k|\pm \mu_5)^2+M_-^2\right ]\Bigg\},
\end{align}
where
\begin{equation}
M_\pm^2\equiv (M\pm a)^2+(\eta\pm\Pi)^2\quad \text{and}\quad \text{Tr}(1)=8NT\sum_n\int\frac{d^3\vec k}{(2\pi)^3}\left [k_0\to \omega^F_n=\frac{(2n+1)\pi}\beta\right ].
\end{equation}
From now on, when we refer to the neutral pion condensate, we will write $\Pi$.
Note that, as explained in appendix \ref{trick}, one is able to write the determinant as the trace of an
operator that is the identity in flavour space in spite of the initial non-trivial flavour structure. 
This facilitates enormously the calculations.

In the search for stable configurations in the potential \eqref{V} we will need the derivatives of the fermion determinant, 
which are basically given by the function $K_1$ that we define as
\begin{equation}\label{trace}
4NK_1=\text{Tr}\sum_\pm\frac 1{(ik_0+\mu)^2-[(|\vec k|\pm\mu_5)^2+M^2]},
\end{equation}
which is clearly divergent in the UV. In this work, we will deal with the NJL model using dimensional regularization (DR) and 
a 3-momentum cut-off ($\Lambda$) both at zero temperature \cite{ikkk2,ikkk3}. The function $K_1$ depending on the regulator 
can be written as follows
\begin{align}\label{K1DR}
\nonumber K_1^{\text{DR}}(M,\mu,\mu_5)=&\frac1{2\pi^2}\Bigg [\Theta(\mu-M)\left \{\mu\sqrt{\mu^2-M^2}+(2\mu_5^2-M^2)\log\left (\frac {\mu+\sqrt{\mu^ 2-M^2}}M\right )\right \}\\
&-\frac12 M^2+\frac1{2}(M^2-2\mu_5^2)\left (\frac1{\epsilon}-\gamma_E + 2 -\log\frac{M^2}{4\pi\mu_R^2}\right )\Bigg ],
\end{align}
\begin{align}\label{K1Lambda}
\nonumber K_1^\Lambda(M,\mu,\mu_5)=&\frac1{2\pi^2}\Bigg [\Theta(\mu-M)\left \{\mu\sqrt{\mu^2-M^2}+(2\mu_5^2-M^2)\log\left (\frac {\mu+\sqrt{\mu^ 2-M^2}}M\right )\right \}\\
&-\frac12 {M^2}+\frac12 (M^2-2\mu_5^2)\log\frac {4\Lambda^2}{M^2}-\Lambda^2\Bigg ].
\end{align}
The quadratically divergent term in the cut-off regularization can be reabsorbed in the couplings $G_{1,2}$. 
After the redefinition, the two results are then identical if we identify
\begin{equation}
\frac1\epsilon-\gamma_E + 2 ~\longleftrightarrow ~\log\frac{\Lambda^2}{\pi\mu_R^2}.
\end{equation}
However in both cases the logarithmic divergence cannot be absorbed \cite{ZJ} unless we include extra terms in 
the Lagrangian like $(\partial\sigma)^2$ and $\sigma^4$. This is of course a manifestation of the non-renormalizability
of the model.
For this reason, we shall assume the scale $\Lambda$ (or equivalently $\mu_R$) to represent a physical cut-off
and write
\begin{align}\label{K1}
\nonumber K_1(M,\mu,\mu_5)=&\frac1{2\pi^2}\Bigg [\Theta(\mu-M)\left \{\mu\sqrt{\mu^2-M^2}+
(2\mu_5^2-M^2)\log\left (\frac {\mu+\sqrt{\mu^ 2-M^2}}M\right )\right \}\\
&-\frac{M^2}2+(M^2-2\mu_5^2)\log\frac{2\Lambda}{M}\Bigg ].
\end{align}
Note that $K_1$ increases with $\mu$ and decreases with $\mu_5$. The derivative of this function will also be used
\begin{align}\label{L1}
\nonumber L_1(M,\mu,\mu_5)\equiv& \frac 1M\frac{\partial K_1}{\partial M}=
-\frac 1{\pi^2}\Bigg[\Theta(\mu-M)\Bigg \{\frac{\mu\mu_5^2}{M^2\sqrt{\mu^2-M^2}}
+\log\left (\frac{\mu+\sqrt{\mu^2-M^2}}M\right)\Bigg \}\\
&+1-\frac{\mu_5^2}{M^2}-\log\frac{2\Lambda}{M}\Bigg].
\end{align}
It verifies the property $L_1(\mu_5=0)>0$.

\section{Search for stable vacuum configurations}
We will now explore the different phases that are allowed by the effective potential \eqref{V} 
by solving the gap equations and analysing the second derivatives to investigate the stable configurations of the different scalar 
and pseudoscalar condensates. The gap equations for the system read
\begin{align}
\nonumber \frac{\sigma}{2G_1}+\sum_{\pm}(M\pm a) K_1^\pm =0, \qquad \frac{\eta}{2G_2}+ \sum_{\pm}(\eta \pm \Pi) K_1^\pm =0\\
\frac{\Pi}{2G_1}+\sum_{\pm}\pm(\eta\pm\Pi) K_1^\pm=0, \qquad \frac{a}{2G_2}+ \sum_{\pm} \pm (M\pm a)K_1^\pm=0
\end{align}
where $K_1^\pm\equiv K_1(M_\pm,\mu,\mu_5)$ (the same convention applies to $L_1$). The second derivatives of the potential are
\begin{gather}
\nonumber V_{\sigma\sigma}=\frac{1}{2G_1}+\sum_{\pm}\left [(M\pm a)^2 L_1^\pm +K_1^\pm \right ], \qquad 
 V_{\eta\eta}=\frac{1}{2G_2}+\sum_{\pm}\left [(\eta \pm \Pi)^2 L_1^\pm +K_1^\pm\right ]\\
\nonumber V_{\pi\pi}=\frac{1}{2G_1}+\sum_{\pm}\left [(\eta \pm \Pi)^2 L_1^\pm +K_1^\pm\right ], \qquad 
V_{aa}=\frac{1}{2G_2}+\sum_{\pm}\left [(M\pm a)^2 L_1^\pm +K_1^\pm\right ]\\
\nonumber V_{\sigma\eta}=V_{\pi a}=\sum_{\pm}(M\pm a)(\eta \pm \Pi)L_1^\pm , \qquad 
V_{\sigma\pi}=V_{\eta a}=\sum_{\pm} \pm (M\pm a)(\eta\pm \Pi) L_1^\pm\\
V_{\sigma a}=\sum_{\pm}\pm \left [(M\pm a)^2 L_1^\pm  + K_1^\pm \right ], \qquad V_{\eta\pi}=
\sum_{\pm}\pm \left [(\eta \pm \Pi)^2 L_1^\pm +K_1^\pm\right ]
\end{gather}
To keep the discussion simple we will assume in the subsequent that $a=0$. However, in Sec. 4 we will see that in a
very tiny region of the parameter space there is evidence of the existence of a phase with $a\neq 0$.

\subsection{Chirally symmetric phase}
We will first consider the phase where none of the fields condenses (in the chiral limit with $m=0$ and $\mu=\mu_5=0$ for
simplicity). 
The gap equations are automatically satisfied, while 
the second derivatives read in this case
\begin{gather}
\nonumber V_{\sigma\sigma}=V_{\pi\pi}=\frac{1}{2G_1}+2K_1, \qquad V_{\eta\eta}=V_{aa}=\frac{1}{2G_2}+2K_1,\\
V_{\sigma\eta}=V_{\sigma\pi}=V_{\sigma a}=V_{\eta\pi}=V_{\eta a}=V_{\pi a}=0.
\end{gather}
After absorbing the quadratic divergence from the cut-off regularization into the coupling constants as mentioned previously
\begin{gather}
\frac{1}{2G_i}-\frac{\Lambda^2}{\pi^2} = \frac{1}{2G_i^r}
\end{gather}
the stability conditions for this phase are $G_{1,2}^r>0$. For simplicity, we will drop the superindex $r$ throughout.

\subsection{Chirally broken phase}\label{CSB}
In this phase we will explore the phase where the field $\sigma$, and only this field, condenses. The gap equations reduce just 
to one
\begin{gather}\label{chiSB}
K_1=-\frac1{4G_1}\left (1-\frac mM\right ).
\end{gather}
Let us first assume $\mu=\mu_5=0$. Then the condition for chiral symmetry breaking (CSB) after absorbing the quadratic divergence
into the coupling constants (or right away in DR for that matter) reads
\begin{gather}\label{genericgap}
M^2\left (\frac{1}2-\log\frac {2\Lambda}M\right )=\frac{\pi^2}{2G_1}\left (1-\frac mM\right ).
\end{gather}

\begin{figure}[h!]
\centering
\includegraphics[scale=0.31]{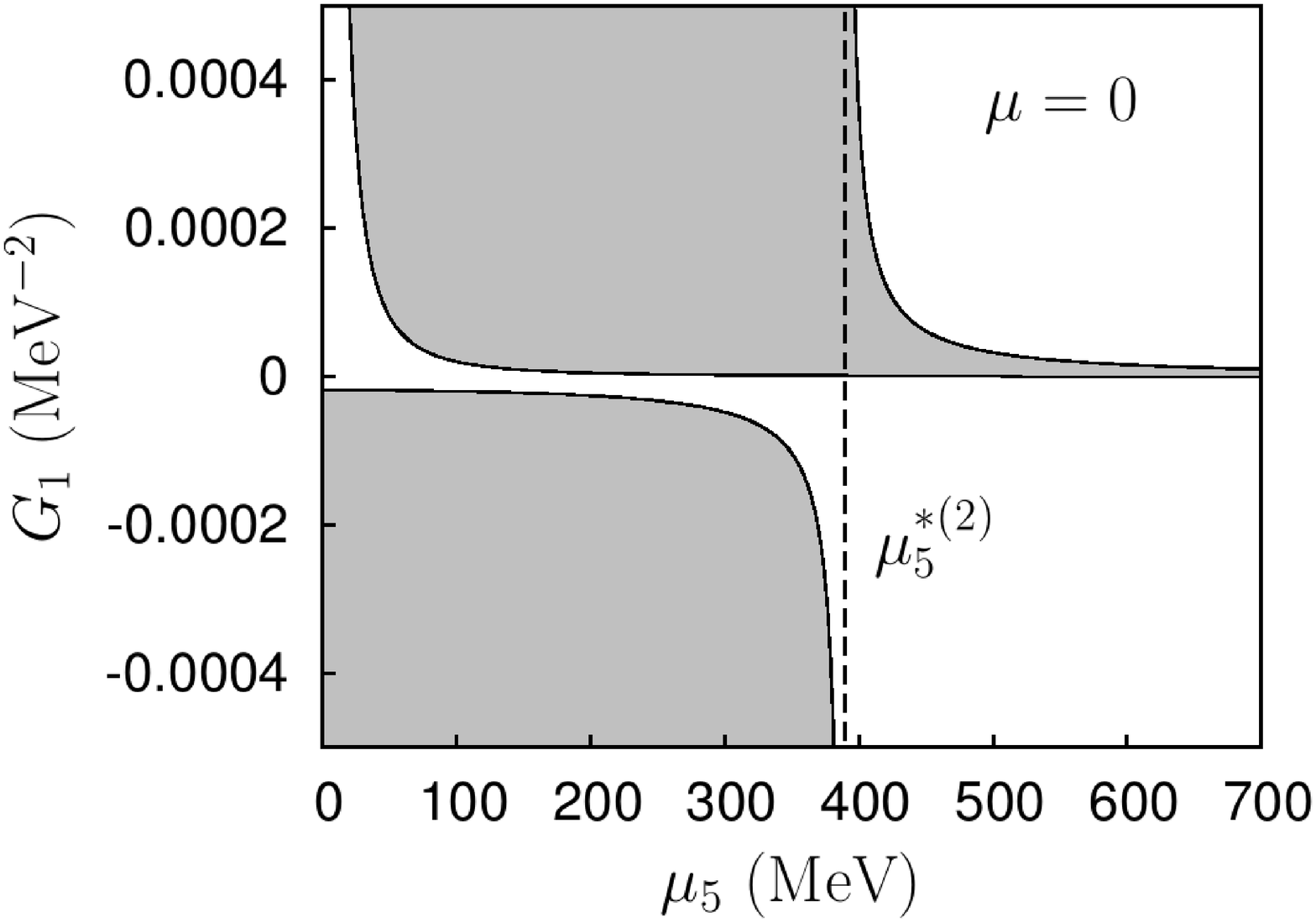}\quad\includegraphics[scale=0.31]{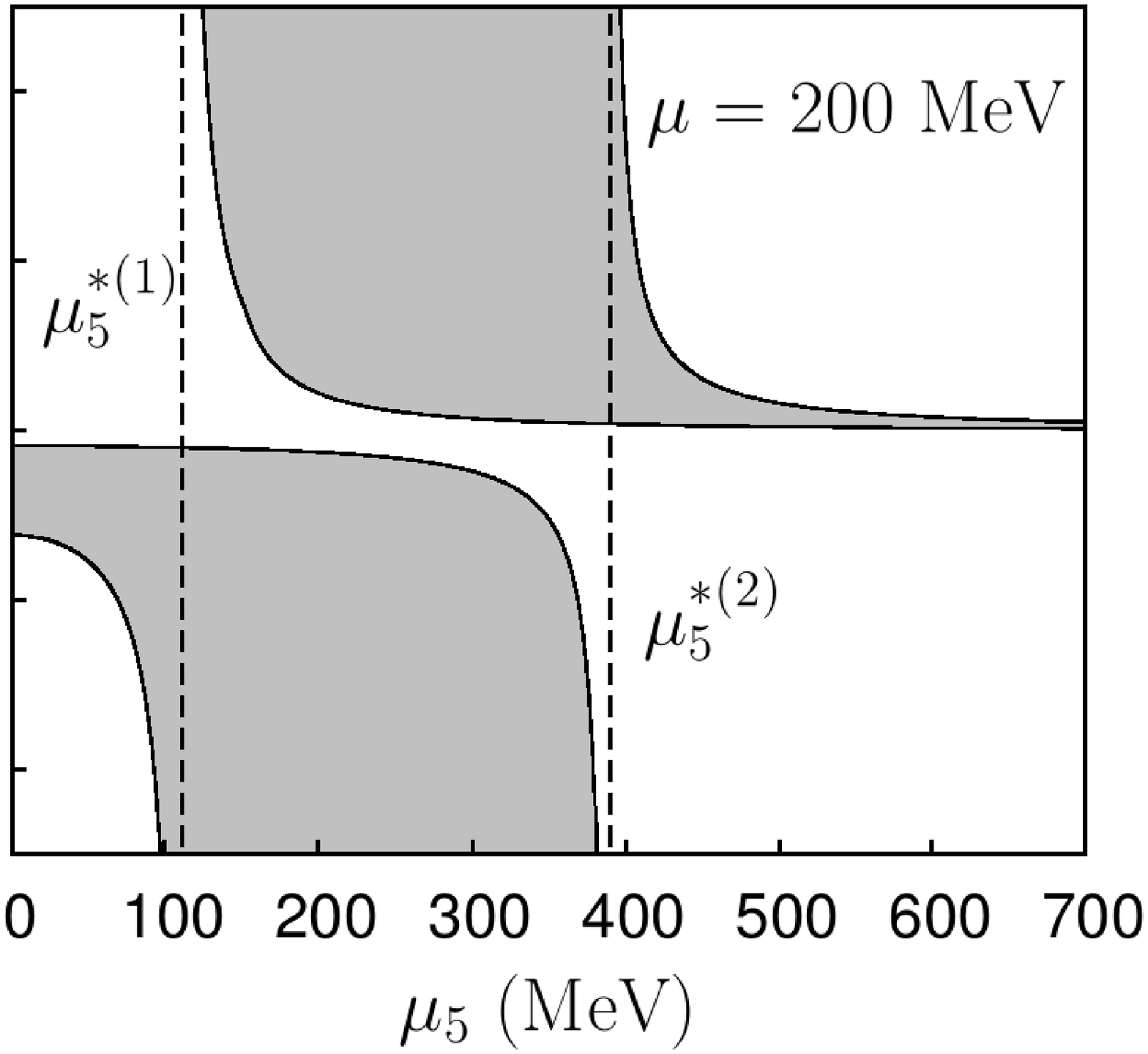}
\caption{Allowed region of $G_1$ as a function of $\mu_5$ with fixed $\mu$ for a stable CSB phase (dark region). The left panel shows $\mu=0$ while the right one corresponds to $\mu=200$ MeV. The figure corresponds to $m=0$ and $\Lambda=1$ GeV.}\label{CSB}
\end{figure}
In Fig. \ref{CSB} we show the region of $G_1$ that provides a stable CSB phase with $m=0$ for non-trivial values of the external drivers. All dimensional quantities scale with $\Lambda$, which we take to be $\Lambda=1$ GeV throughout.
Two discontinuities appear in the plot. The first one is found at
\[\left (\mu_5^{*(1)}\right )^2=\frac{\mu^2}2\left [\Theta(\mu-\mu^*)\left (1-\frac1{2\ln\frac{2\Lambda}\mu}\right )+\Theta(\mu^*-\mu)\frac1{\ln\frac\Lambda\mu}\right ]\]
with
\[\mu^*\equiv\exp\left [-\frac14\left (3-2\ln2+\sqrt{9+4\ln2+4\ln^22}\right )\right ]\Lambda\approx 0.265\Lambda,\]
while the second one can be written analytically only if $\mu<2\exp[-\frac14(1+\sqrt5)]\Lambda\approx 0.891\Lambda$. In this case, the second discontinuity is given by
\[\left (\mu_5^{*(2)}\right )^2=(3-\sqrt 5)\Lambda^2\exp\left [-\frac12\left (1+\sqrt 5\right )\right ]\approx (0.389\Lambda)^2.\]
For $\mu=0$ and $\mu=200$ MeV, the condition $\mu<0.891\Lambda$ is satisfied and the previous equation can be used to find the discontinuity, which is clearly independent of $\mu$. 
The limit $\mu\to 0$ reduces to $G_1<0$, a known result from a previous work on the NJL model in DR \cite{ikkk3}. Finally, note that the restriction for $G_2$ is simply $\frac{1}{G_2}>\frac{1}{G_1}$.

The meson spectrum for any value of the external chemical potentials is given by the second derivatives at the local minimum
\[\nonumber V_{\sigma\sigma}=\frac{m}{2G_1M}+2M^2L_1, \qquad V_{\eta\eta}=\frac{m}{2G_1M}+\frac12\left (\frac{1}{G_2}-\frac{1}{G_1}\right )\]
\[\nonumber V_{\pi\pi}=\frac{m}{2G_1M}, \qquad V_{aa}=\frac{m}{2G_1M}+\frac12\left (\frac{1}{G_2}-\frac{1}{G_1}\right )+2M^2L_1\]
\begin{equation}\label{spectr}
V_{\sigma\eta}=V_{\sigma\pi}=V_{\sigma a}=V_{\eta\pi}=V_{\eta a}=V_{\pi a}=0,
\end{equation}
where one has to use a bare quark mass $m$ of the same sign as the coupling $G_1$ so as to provide a positive pion mass. 

The stability conditions read
\[\frac1{G_2}>\frac1{G_1}\left (1-\frac mM\right ),\qquad 2M^2L_1>\max\left[-V_{\pi\pi},-V_{\eta\eta}\right ].\]
Let us set once again $\mu=\mu_5=0$. Then $L_1>0$ and the second covexity condition is always met if the first one is fulfilled.
In this case the mass spectrum obeys the relation
\[m_\sigma^2-m_\pi^2=m_a^2-m_\eta^2>0\]
in analogy to the situation in QCD. In addition the following relation also holds
\[m_a^2-m_\sigma^2=m_\eta^2-m_\pi^2,\]
and the difference $m_\eta^2-m_\pi^2$ is positive (like the analogous one in QCD \cite{vw}) 
provided that  $\frac{1}{G_2}-\frac{1}{G_1}>0$.

\begin{figure}[h!]
\centering
\includegraphics[scale=0.3]{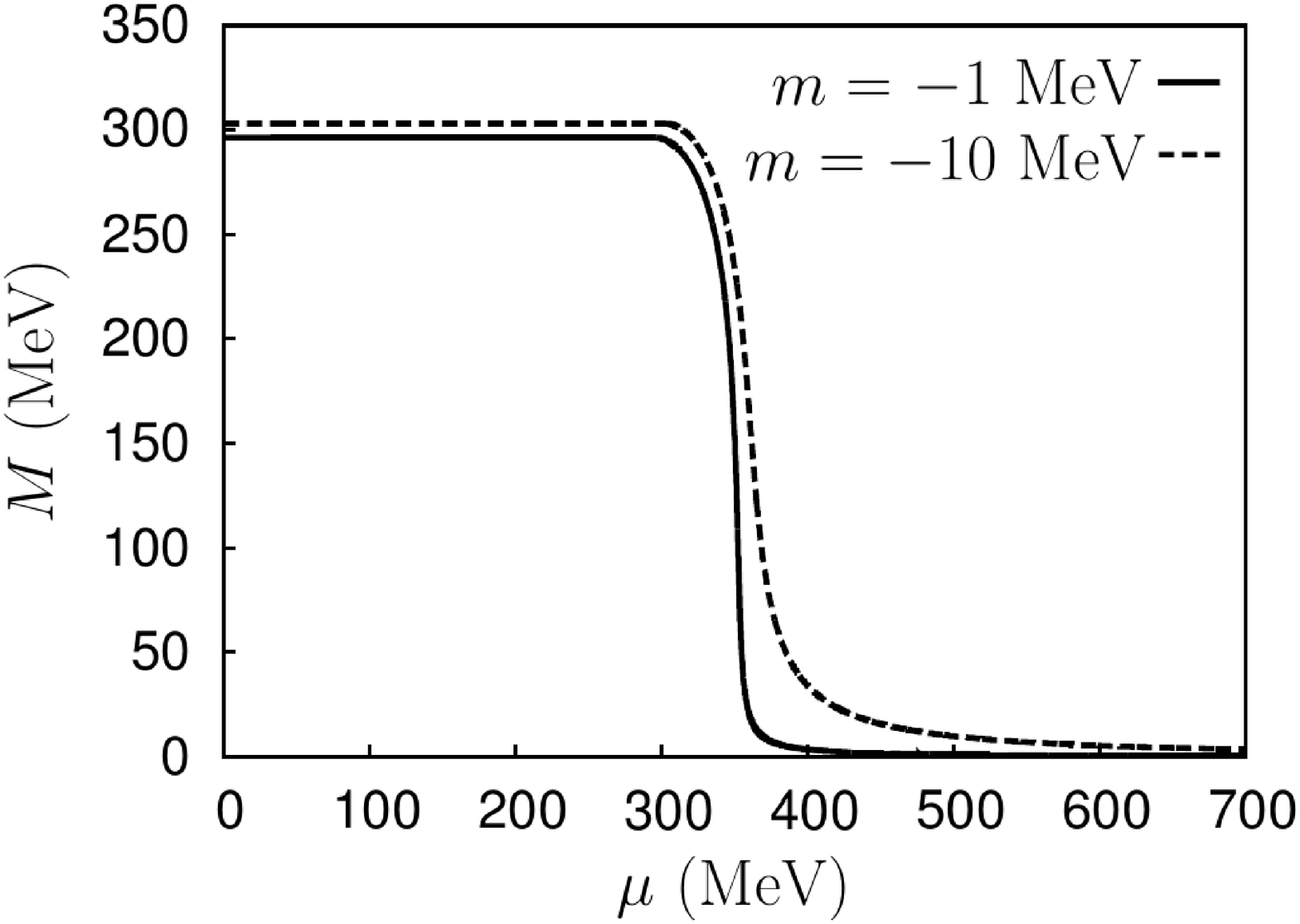}\quad \includegraphics[scale=0.3]{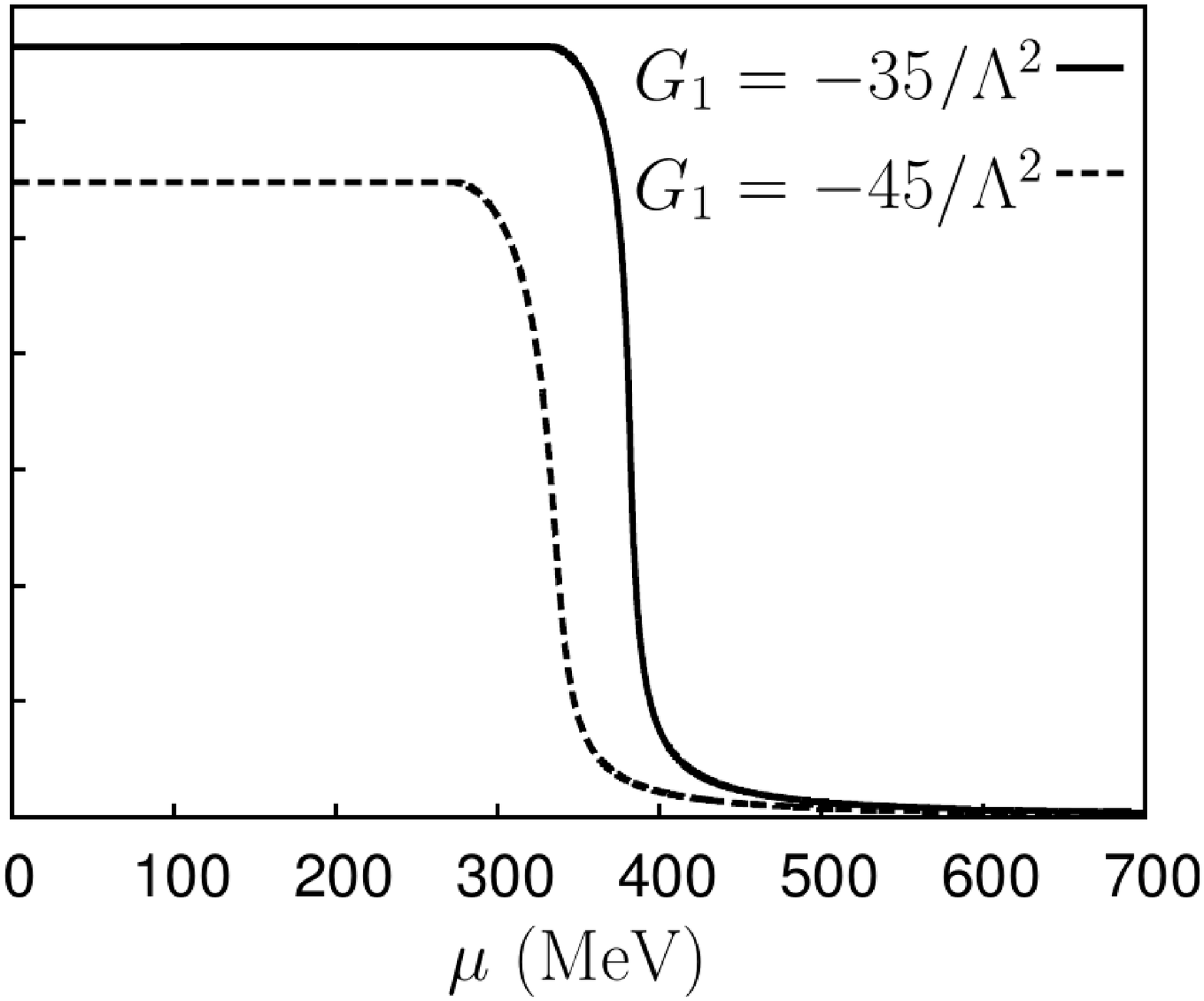}
\caption{Evolution of the constituent quark mass $M$ depending on $\mu$. For both plots we set $G_2=-45/\Lambda^2$ 
with $\Lambda=1$ GeV and $\mu_5=0$. In the left panel, we fixed $G_1=-40/\Lambda^2$ and plot for different values of $m$. 
In the right panel instead, we fixed $m=-5$ 
MeV in order to examine the variation of $G_1$. The transition becomes sharper as $m$ decreases.}\label{japos1}
\end{figure}

Let us now examine in detail the dependence of the chiral condensate on the external chemical potentials.
In Figure \ref{japos1} we present the evolution of the constituent quark mass as a function of the vector chemical potential 
for different values of the current quark mass and coupling $G_1$ (left and right panels respectively) with $\mu_5=0$. 
Both the bare quark mass and the coupling $G_1$ are taken to be negative, as just explained above. There is chiral restoration 
around a certain value of the chemical potential that depends mostly on $G_1$; this
phenomenon of chiral restoration is well known in the NJL model \cite{buba} and it is possibly the
main reason that this simple model fails to reproduce correctly the transition
to nuclear matter. The transition becomes sharper as the value $m=0$ is approached.

\begin{figure}[h!]
\centering
\includegraphics[scale=0.45]{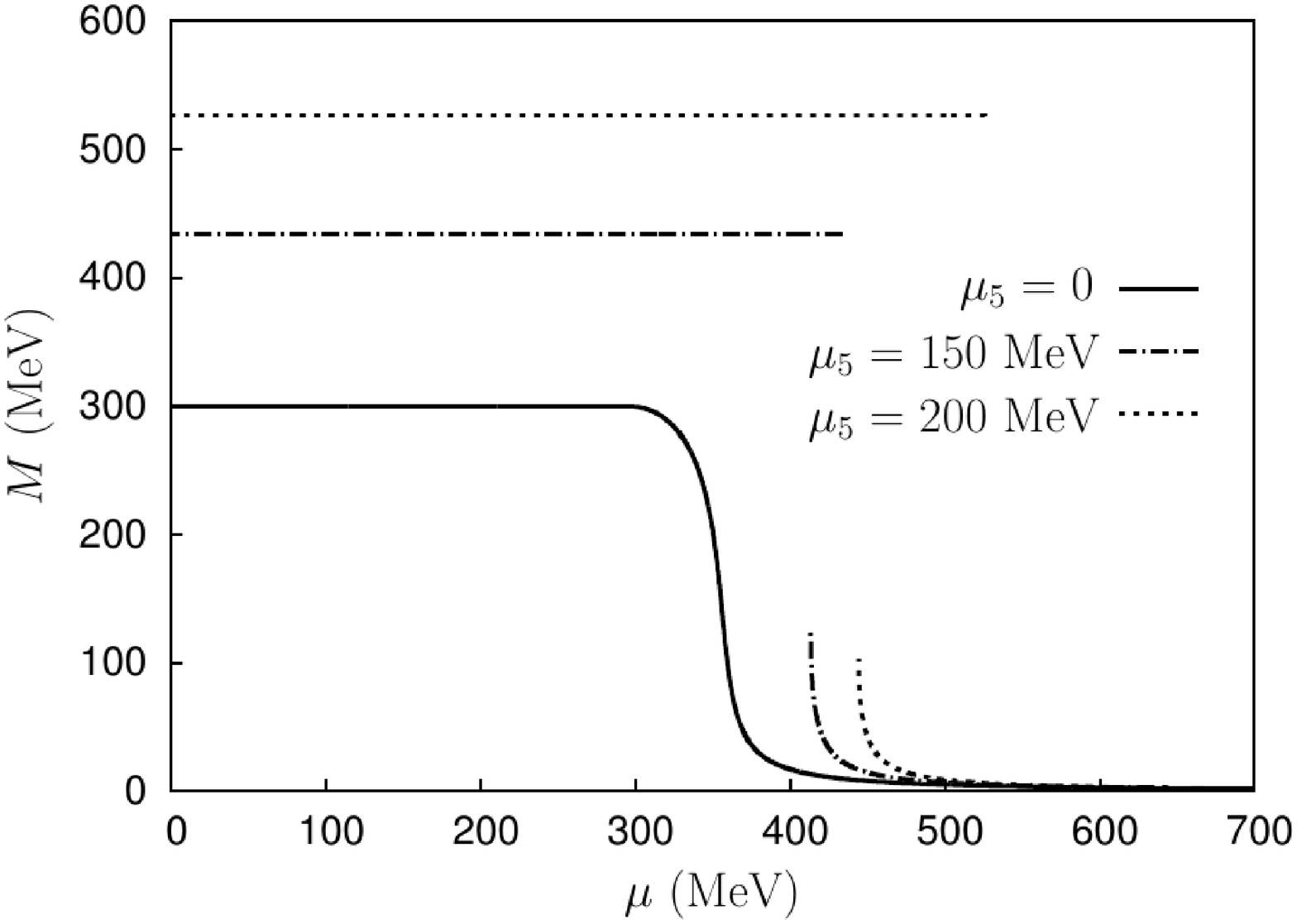}
\caption{Evolution of the constituent quark mass $M$ depending on $\mu$ for different values of the axial chemical potential $\mu_5$ setting $m=-5$ MeV, $G_1=-40/\Lambda^2$ and $G_2=-45/\Lambda^2$. The drawn lines correspond to locally stable phases and accordingly the absence of a continuous line in the cases where $\mu_5\neq0$ is due to the fact that the Hessian matrix is not positive definite. The transition to a chirally restored phase changes to a first order one as $\mu_5$ increases.}\label{japos2}
\end{figure}

In Fig. \ref{japos2} we observe the influence of the axial chemical potential $\mu_5$ on the restoration of chiral symmetry
that always takes place in the NJL as $\mu$ increases. For high values of the axial 
chemical potential, the plateau appearing for $M>\mu$ acquires bigger values and spreads over a wider range of $\mu$. 
At some point, the solution of the gap equation shows a stable and a metastable solution that must necessarily flip thus implying 
a jump of the constituent quark mass at some value of the chemical potential where both solutions coexist. Between these solutions, another unstable solution exists, but is not shown in the plot since the Hessian matrix is not positive definite. The jump represents 
a first order phase transition from $\mu<M$ (=constant) to a non-constant $M$ smaller than the chemical potential.

\begin{figure}[h!]
\centering
\includegraphics[scale=0.3]{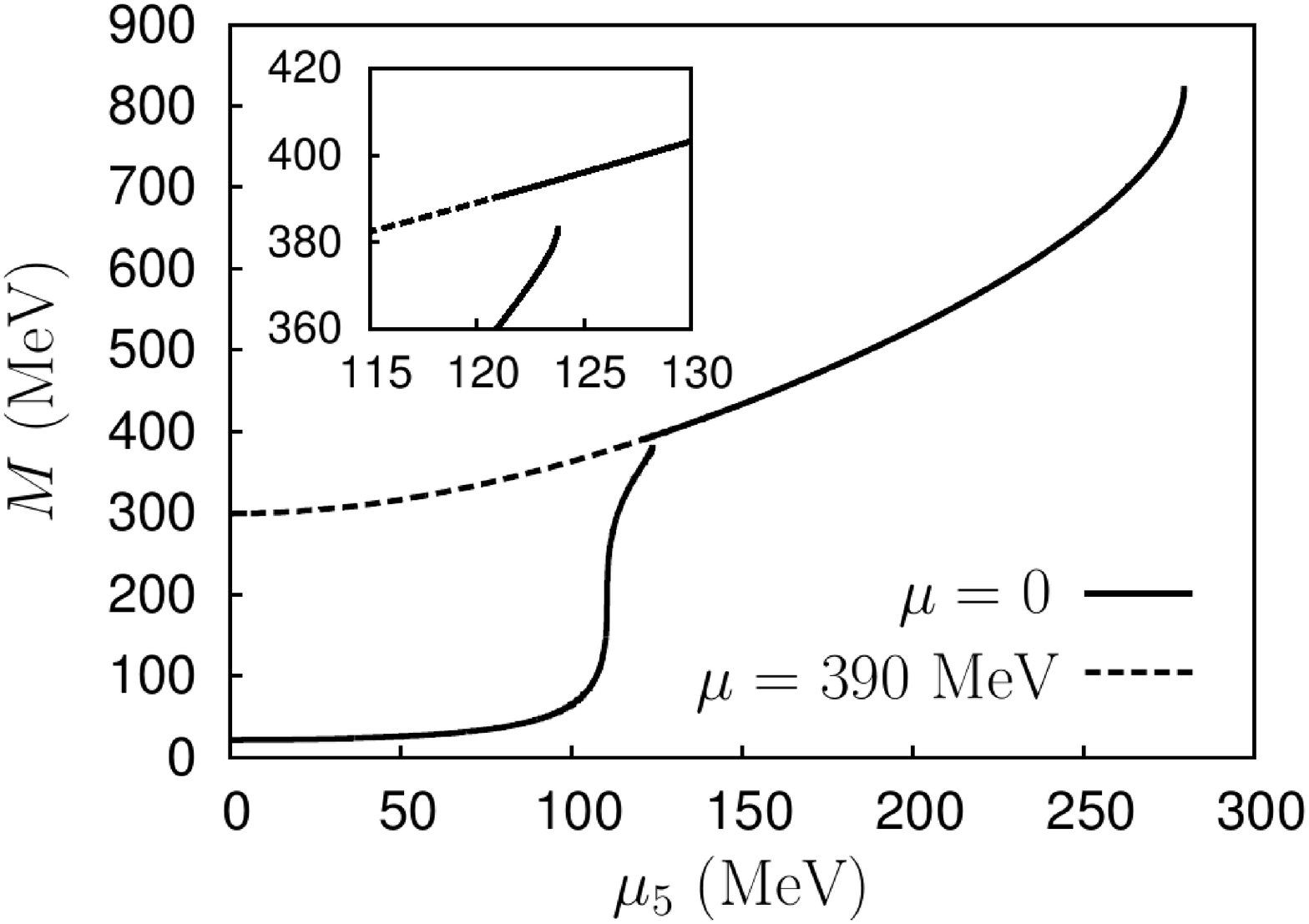} \quad \includegraphics[scale=0.3]{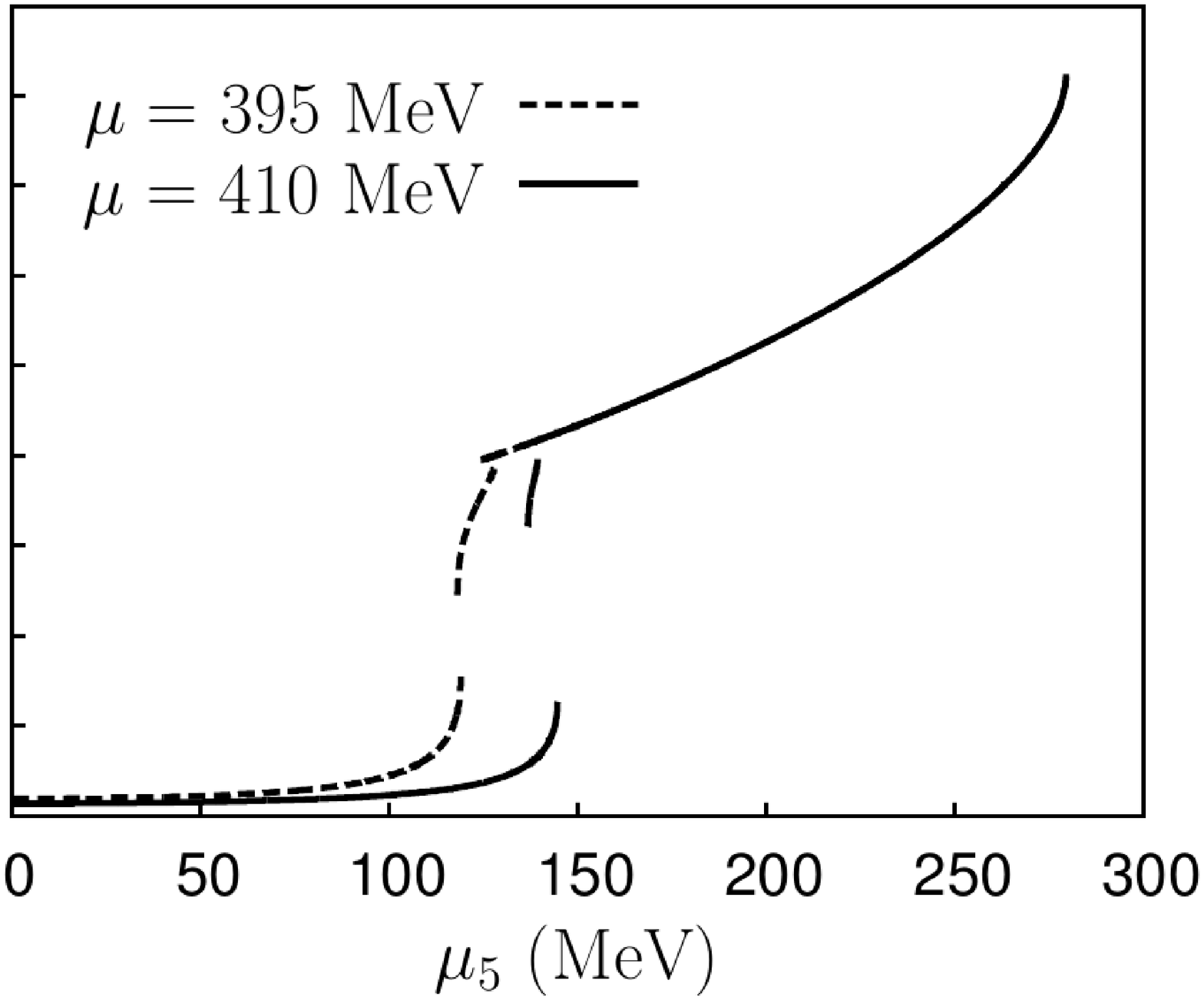}
\caption{Evolution of the constituent quark mass $M$ depending on $\mu_5$ for different values of the chemical potential $\mu$ setting $m=-5$ MeV, $G_1=-40/\Lambda^2$ and $G_2=-45/\Lambda^2$. Both graphics show the regions where all the second derivatives are positive.
Certain values of $\mu_5$ exhibit coexisting solutions implying first order phase transitions. In the left panel, we show a plot for $\mu=0$ (or indeed for any $\mu<M$) and $\mu=390$ MeV. The second curve exhibits a small jump that is shown more detailed in the inset. The right panel corresponds to $\mu=395$ (two jumps) and 410 MeV (probably only one jump). This plot shows that the NJL with external drivers has a rather complex phase diagram.}\label{japos3}
\end{figure}

It may be helpful to show a plot of the same constituent quark mass depending on $\mu_5$ for different values of $\mu$. In the left panel of Fig. \ref{japos3} we display such evolution for $\mu=0$ and 390 MeV. The first curve is valid for any $\mu<M\approx 300$ MeV while the second one shows a small discontinuity that represents a first order phase transition within the CSB phase. A detail of the jump is presented in the inset. Note that both curves coincide after the jump and stop at $\mu_5\sim 280$ MeV since beyond this value, the phase becomes unstable, as presented previously in Fig. \ref{CSB}.

In the right panel, we present the values of $\mu=395$ and 410 MeV, which correspond to qualitatively different cases. The curve for $\mu=395$ MeV shows two separate regions where the function is bivaluated. First, the lower and intermediate branches share some common values of $\mu_5$ even that it cannot be appreciated in the plot. Thus, a first order phase transition must take place within this region. The same behaviour happens for the intermediate and the upper branches, implying another first order phase transition. For bigger values of $\mu_5$ one recovers the tendency of $\mu=0$ as in the previous case. The curve $\mu=410$ MeV is somewhat similar to the previous one but now with a trivaluated region:   for a certain small range of $\mu_5$ the three branches may be reached and therefore one or two jumps may take place. For bigger values of $\mu$, the intermediate branch disappears and only one jump may take place.

All the jumps in Fig. \ref{japos3} are due to the presence of unstable regions that would connect the different branches of the same curve. Here, it can be shown that $V_{\sigma\sigma}<0$ is the responsible for these unstable zones. On the other hand, $V_{aa}$ is simply $V_{\sigma\sigma}$ with a positive shift and the restriction $V_{aa}>0$ does not add anything new.

We want to stress that all the first order phase transitions just explained are a direct consequence of the addition of $\mu_5$ to the problem. No other assumptions are made beyond using the mean field approximation.

\section{Isosinglet pseudoscalar condensation and parity-breaking}
Next we focus in the analysis of parity violating phases. It turns out that the only stable one corresponds to
condensation in the isoscalar channel. Neutral pseudoscalar isotriplet condensation, either with or without CSB, does not lead to a stable
termodynamical phase\footnote{This is at variance with the QCD- inspired effective theory analysis of \cite{anesp} where the
possibility of a condensation in the isotriplet channel was proven.}. Now, in addition 
to the scalar condensate $\sigma$ that was explored in the previous section we will allow for a non-vanishing 
isosinglet pseudoscalar condensate $\eta$. The gap equations now turn to be
\begin{gather}\label{PB}
M=\frac m{G_1}\frac1{\frac1{G_1}-\frac1{G_2}}, \qquad K_1=-\frac{1}{4G_2}.
\end{gather}
The first gap equation shows that the scalar condensate exhibits a remarkable independence on the external chemical potentials
as it turns out to be constant once the parameters of the model are fixed. Unlikely the $\eta$ condensate does depend 
on the external drivers through the second equation. Moreover, from the first equation one finds that 
in the parity breaking phase $m=0$ iff $G_1=G_2$; namely, the parity breaking $\eta$ condensate is a stationary  
point of the effective potential \eqref{V} only when the chiral and $U(1)_A$ symmetries 
are explicitly preserved or broken at the same time in the NJL Lagrangian \eqref{lagNJL}. However this
stationary point would not be a true minimum but a stationary point with two flat directions. 
The more general case where $m\neq 0$ and $G_1\neq G_2$ is thus the only possibility to have 
a genuine parity breaking phase. We will see in a moment how as one takes the limit $m\to0$, the narrow window to have access to this 
phase disappears.

The second derivatives read
\begin{gather}
\nonumber V_{\sigma\sigma}=\frac{1}2\left (\frac{1}{G_1}-\frac{1}{G_2}\right )+2M^2L_1, 
\qquad V_{\eta\eta}=2\eta^2L_1, \qquad V_{\sigma\eta}=2M\eta L_1\\
\nonumber V_{\pi\pi}=\frac{1}2\left (\frac{1}{G_1}-\frac{1}{G_2}\right )+2\eta^2L_1, 
\qquad V_{aa}=2M^2L_1, \qquad V_{\pi a}=2M\eta L_1\\
\nonumber V_{\sigma\pi}=V_{\sigma a}=V_{\eta\pi}=V_{\eta a}=0.
\end{gather}
We find that the Hessian matrix is not diagonal but has a block structure with two isolated sectors $\sigma-\eta$ 
and $\pi-a$ that reflect the mixing of states with different parity \cite{anesp,aep}. The determinants of these blocks are
\[\det(V^{\sigma,\eta})=\eta^2 L_1\left (\frac{1}{G_1}-\frac{1}{G_2}\right ), \qquad  
\det(V^{\pi, a})=M^2L_1\left (\frac{1}{G_1}-\frac{1}{G_2}\right ),\]
and thus, the resulting conditions for this phase to be stable reduce to
\begin{equation}\label{conditions}
L_1>0, \qquad \left (\frac{1}{G_1}-\frac{1}{G_2}\right )>0.
\end{equation}

The second of the previous conditions leads to a peculiar ordering of the physical meson spectrum. 
Recall that in the chiral symmetry breaking phase we had
\[m_a^2-m_\sigma^2=m_\eta^2-m_\pi^2=-\frac N2\left (\frac{1}{G_1}-\frac{1}{G_2}\right ),\]
and therefore, a stable parity breaking phase is not compatible with a fit to the phenomenology. 
Thus parity breaking in the NJL model corresponds to a choice of parameters that makes this
model quite different from QCD predictions \cite{vw}. In other words, the NJL model with a stable
parity breaking phase will have nothing to do with QCD.
Note that the above differences are independent
of the phase in which the theory is realized (that is, they are independent of $\mu,\mu_5$).

The rest of the possible phases with a vanishing $a$ require $m=0$ to satisfy the gap equations; they are not true minima. 
In particular, there is no phase with parity breaking and $\sigma=0$.

\subsection{Transition to the parity breaking phase}
In this section we will analyse the characteristics of the transition to the phase where parity is broken.
First of all, let us define $M_0$ as the solution to $M_0=M(G_1,\mu=\mu_5=0)$ in the CSB phase given by Eq. \eqref{chiSB}. 
Recall the inequality $V_{\eta\eta}>0$ of the same phase given in Eq. \eqref{spectr} and the stability condition 
of the parity breaking phase in Eq. \eqref{conditions}. Putting all of them together yields the following inequalities
\[0<\frac1{G_1}-\frac1{G_2}<\frac m{G_1M_0}.\]
The second inequality can be inserted in the first gap equation of the parity breaking phase (see Eq. \eqref{PB}) 
to show that in this phase, $M>M_0$. The same set of inequalities can be rewritten as
\begin{equation}\label{window}
\frac1{G_1}\left (1-\frac{m}{M_0}\right )<\frac1{G_2}<\frac1{G_1},
\end{equation}
which means that $G_1$ and $G_2$ necessarily have the same sign, while in the CSB phase $G_2$ had no restriction and could
 have opposite sign. 
This set of inequalities represents the necessary condition to have a transition from the CSB to a parity breaking phase, 
as they provide the stability conditions of both phases. Notice that the model allows a narrow window of $G_2$ 
(once $G_1$ is fixed) so that both phases may take place depending on the value of the external drivers. 
In the limit $m\to0$, this window closes and no parity breaking can be found. 

Let us recall the gap equation in the CSB phase Eq. \eqref{chiSB} and assume $\mu=\mu_5=0$. 
Provided that Eq. \eqref{window} is satisfied, it follows that
\[K_1=-\frac1{4G_1}\left (1-\frac{m}{M_0}\right )>-\frac1{4G_2}.\]
In the parity breaking phase, the gap equation is $K_1=-\frac1{4G_2}$; therefore to get
into this phase from the familiar CSB one, $K_1$ has to decrease, i.e. from (\ref{chiSB}) we see that $M$ must increase, 
$M(\mu,\mu_5)>M_0$. Let us point out the fact that the condition $L_1>0$ from the parity broken phase 
is stronger than the one from the CSB one so the former will remain to provide stability to both phases. 
Let us describe how this process takes place first for $\mu=0$ and finally for $\mu\neq 0$.

\subsection{Phase transition with $\mu=0$}
Let us simplify the analysis by setting $\mu=0$ and let us study the dependence on $\mu_5$, which makes $M$ increase from its  initial value $M_0$. At some critical value such that
\begin{equation}
M^c\equiv M(\mu_5^c)=\frac m{G_1}\frac1{\frac1{G_1}-\frac1{G_2}},
\end{equation}
where the critical value of the axial chemical potential is
\[(\mu_5^c)^2=\frac{M_c^2}2-\frac1{4\log \frac {2\Lambda}{M_c}}\left (M_c^2-\frac{\pi^2}{G_2}\right ),\]
$m_\eta$ vanishes, and from here on, we get into the parity breaking phase via a 2nd order phase transition, 
where $M$ remains frozen as discussed while the dependence on $\mu_5$ is absorbed into a non-vanishing $\eta$ condensate. 
The dependence of $K_1$ on  $M_\pm^2$ will be now on $M_c^2+\eta^2$. Note that $(\mu_5^c)^2>0$ and therefore, 
a threshold in $M^c$ follows.

\begin{figure}[h!]
\centering
\includegraphics[scale=0.45]{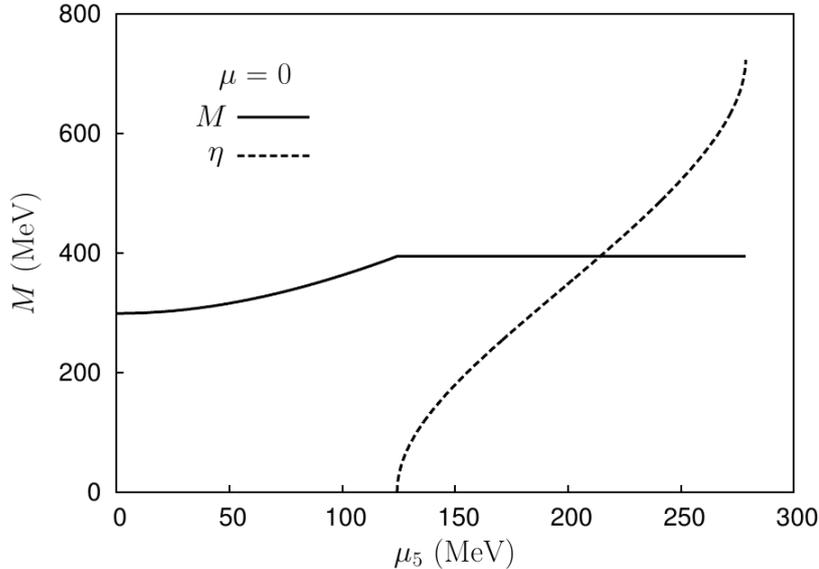}
\caption{$M$ and $\eta$ dependence on $\mu_5$ for $\mu<M_0$, $G_1=-40/\Lambda^2$, $G_2=-39.5/\Lambda^2$, $m=-5$ MeV and $\Lambda=1$ GeV.}\label{M,eta(mu5,mu=0)}
\end{figure}

In Fig. \ref{M,eta(mu5,mu=0)} we present a plot showing the evolution of $M$ and $\eta$ with respect to $\mu_5$ for 
$\mu=0$ (or any $\mu<M_0\approx 300$ MeV). In the CSB phase $M$ grows with $\mu_5$ up to the critical value $M^c$, 
the point where this magnitude freezes out, and $\eta$ acquires non-trivial values, also growing with the 
axial chemical potential. At $\mu_5\simeq 0.28\Lambda$, this phase shows an endpoint and beyond, no stable solution exists. 
This point is the same one that we found in the CSB phase, meaning that the model becomes unstable at such 
value of $\mu_5$, no matter which phase one is exploring.

\subsection{Phase transition with $\mu>0$} 
\begin{figure}[h!]
\centering
\includegraphics[scale=0.3]{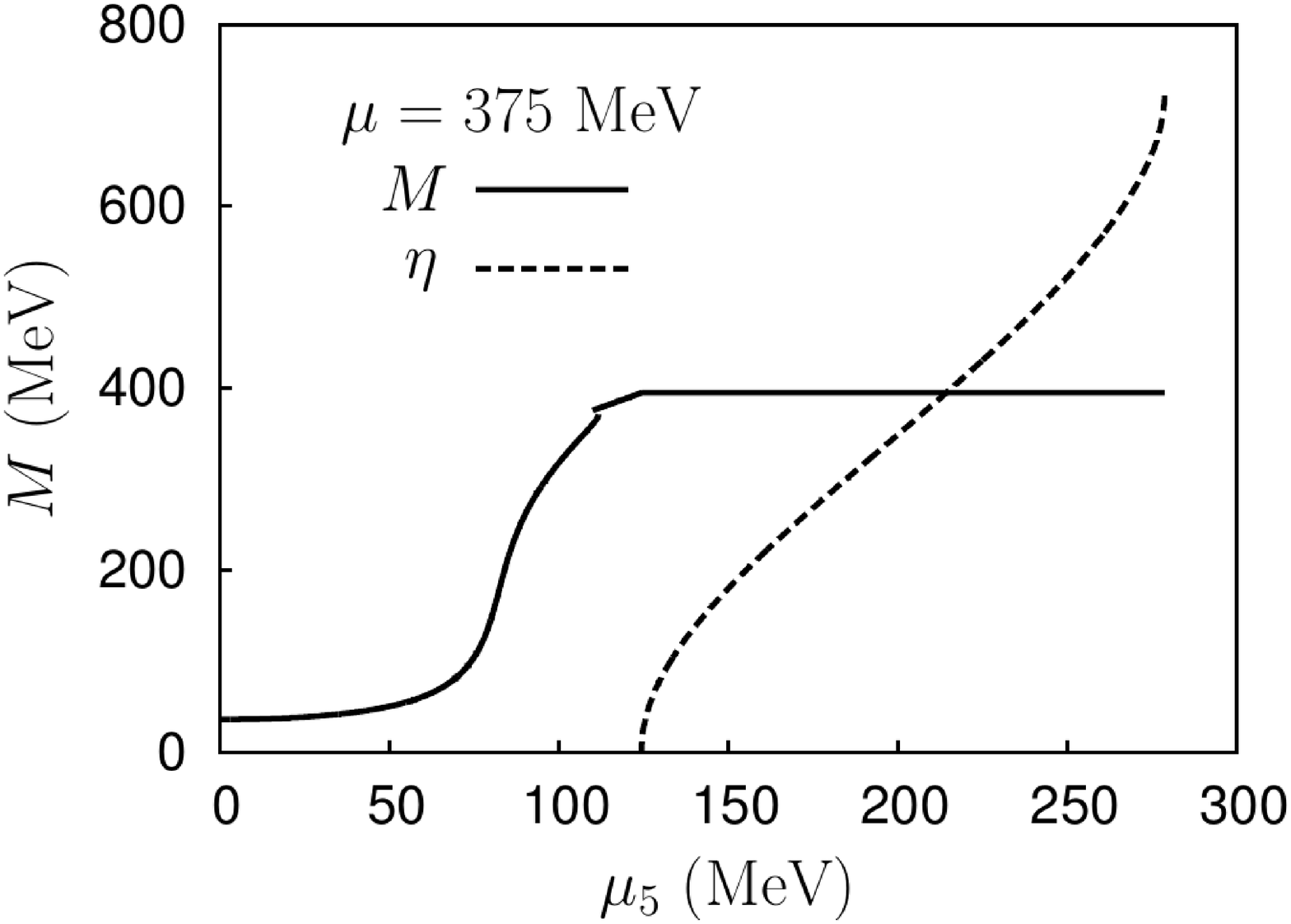}\quad \includegraphics[scale=0.3]{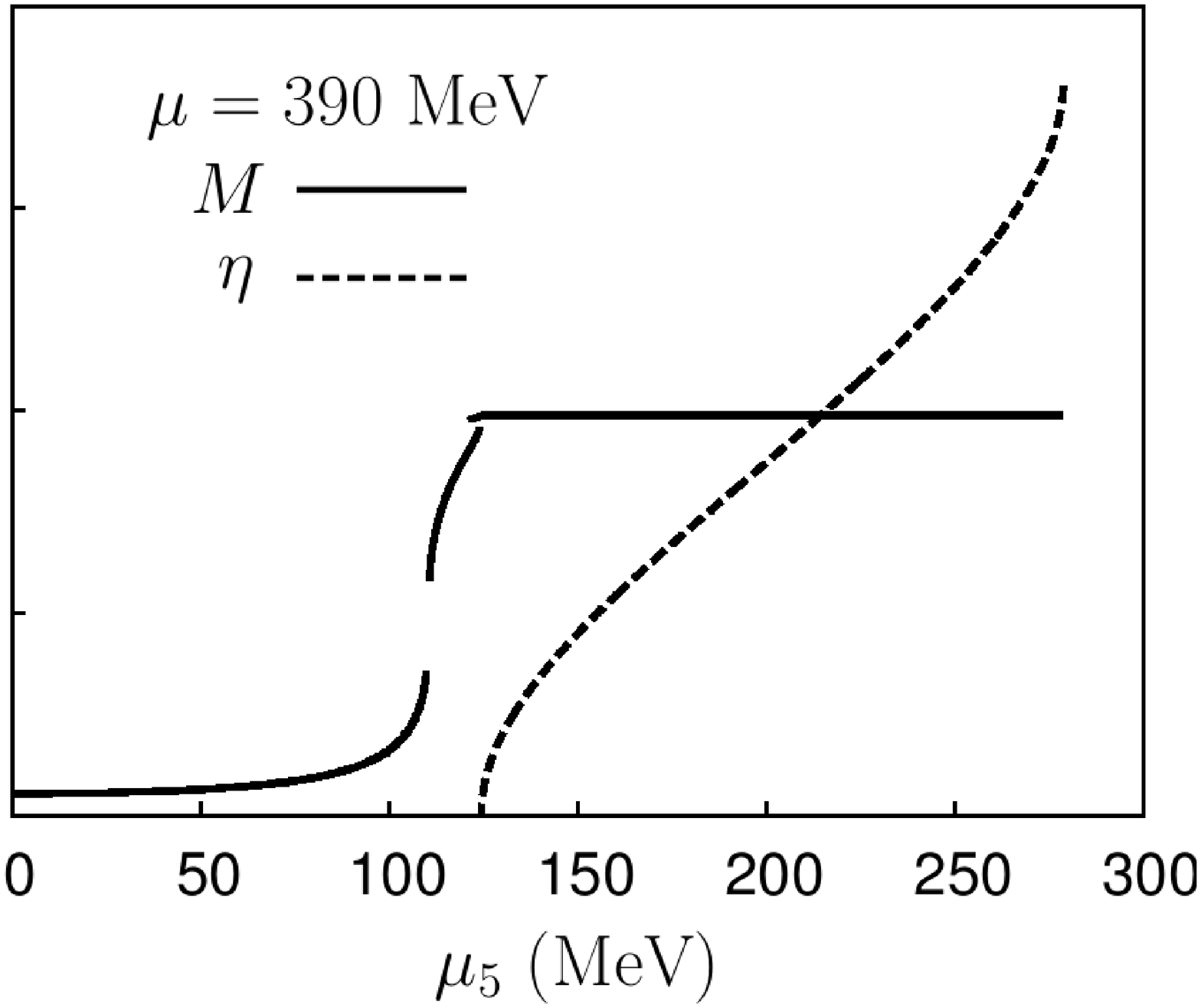}
\includegraphics[scale=0.3]{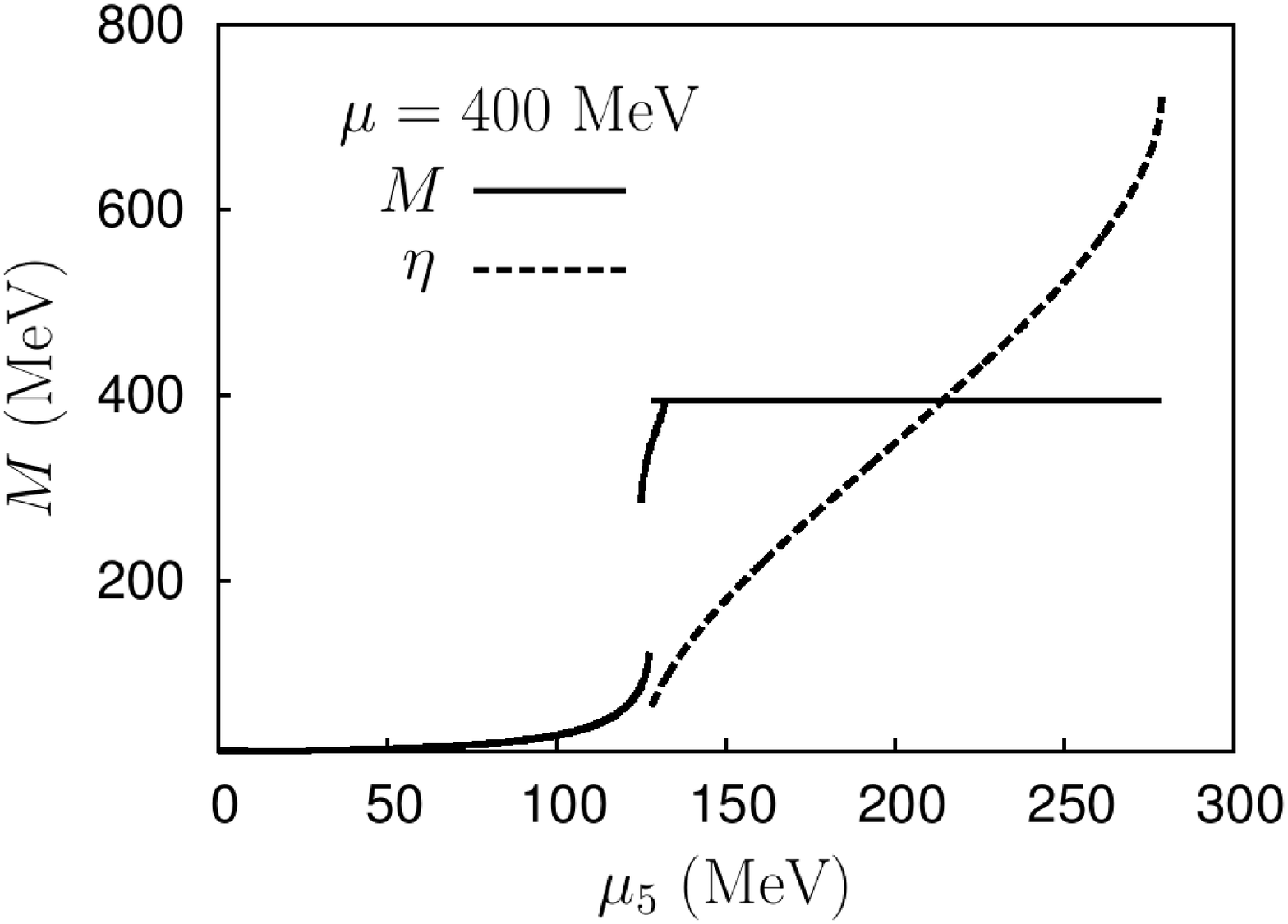}\quad \includegraphics[scale=0.3]{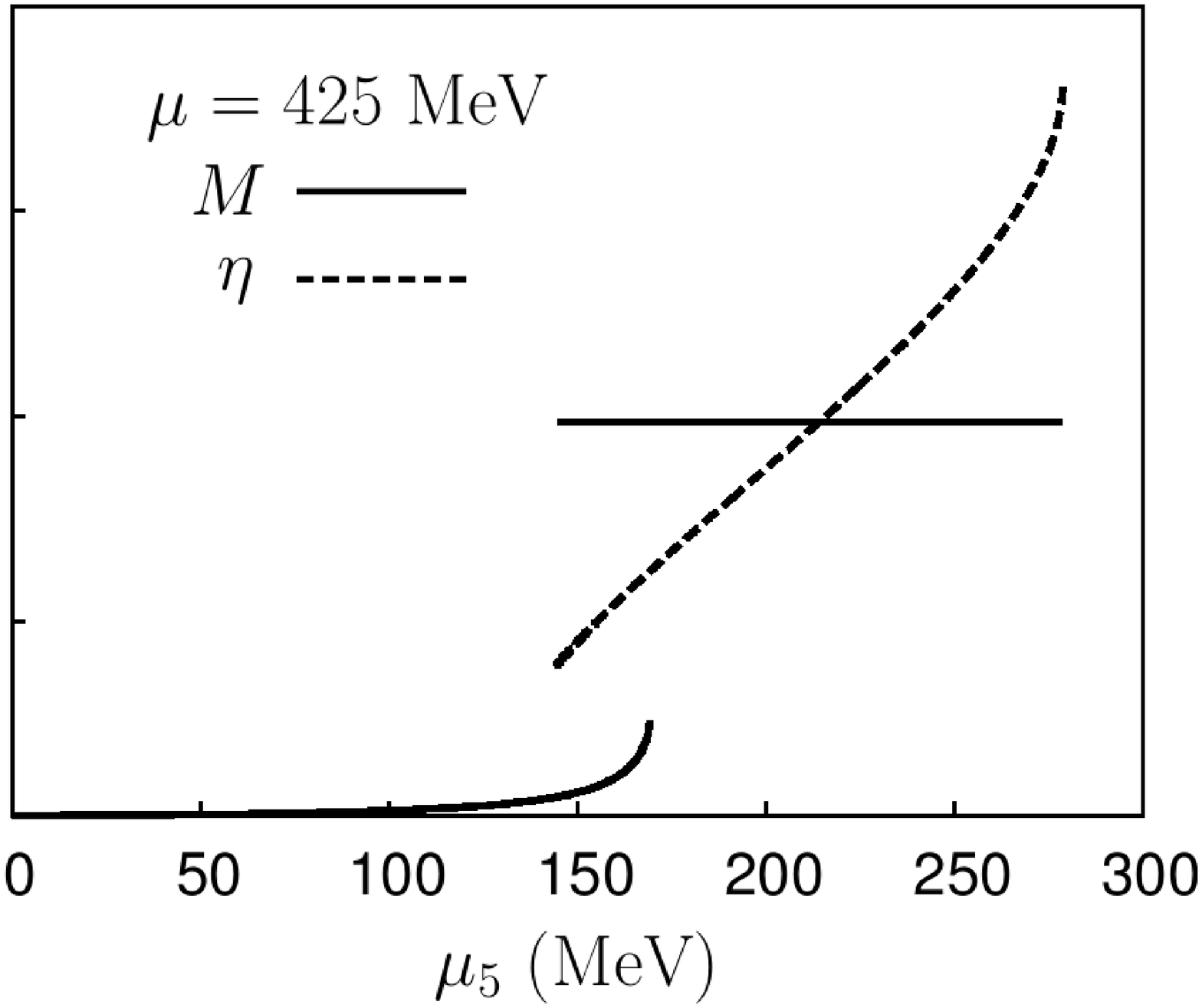}
\caption{$M$ and $\eta$ dependence on $\mu_5$ for $\mu=375$, 390, 400 and 425 MeV, $G_1=-40/\Lambda^2$, $G_2=-39.5/\Lambda^2$, $m=-5$ MeV and $\Lambda=1$ GeV. The graphics show the regions where all the second derivatives are positive.
Certain values of $\mu_5$ exhibit coexisting solutions implying first order phase transitions.
The first jump in the plot for $\mu=390$ MeV shows a very small region of $\mu_5$ where the function is not defined.
This region is characterised by $V_{aa}<0$, thus suggesting a phase with a non-trivial scalar isotriplet condensate.
This is the only region where we have found indications for a phase with $a\neq 0$.
The landscape of first order phase transitions in the constituent quark mass is essentially the same as the one explained in Fig. \ref{japos3}.}\label{mu-evolut}
\end{figure}

The presence of both chemical potentials makes the function $K_1$ exhibit more complicated features. 
As $K_1$ decreases with $\mu_5$ and $\mu$ does the opposite job, $\mu_5$ needs larger values 
than $\mu$ to reach the parity breaking phase. In Fig. \ref{mu-evolut}, we present a set of plots with 
the evolution of both $M$ and $\eta$ for non-vanishing values of the chemical potential. As before we take the value 
$\Lambda=1$ GeV to make the model in order to have some QCD-inspired intuition. Of course everything scales with $\Lambda$.

In the upper panels, we set $\mu=375$ MeV (left) and $\mu=390$ MeV (right), both of them $M_0<\mu<M^c$, where jumps in $M$ 
are observed in the parity even phase together with tiny metastable regions. 
This behaviour is very similar to the one described in Fig. \ref{japos3} with the subtlety that we inverted 
the sign of $\frac1{G_1}-\frac1{G_2}$ and therefore, the parity-odd phase may be reached.

In addition, this change of sign shifts the second derivative $V_{aa}$, which is the only responsible for the apparent 
big jump in the $\mu=390$ MeV window (the one with lower $\mu_5$). It should be clear that $L_1>0$ since $M$ is growing 
with $\mu_5$. However, the second derivative $V_{aa}$ becomes negative due to this shift while all the other 
derivatives remain positive. If for a moment we forgot $V_{aa}$, the curve would be smoothly increasing and we 
would only have the other tiny jump close to the flat region of constant $M$.
However, the fact that this second derivative becomes negative leads to a small 
range of $\mu_5$ where no solution exists. Hence, it seems natural to think that the system goes away from the phase 
with $a=0$ and acquires a non-trivial scalar isotriplet condensate. 
We emphasize this region is really tiny and depends crucially on the specific values for the parameters, even disappearing for $G_1>-30/\Lambda^2$.
Both graphics show a smooth transition to the parity-odd phase, say, via a 2nd order phase transition 
with the same characteristics of the previous section with $\mu=0$.

On the other side, in the lower panels, we set $\mu=400$ MeV (left) and $\mu=425$ MeV (right) with $\mu>M^c$ and observe 
what we could more or less expect from Fig. \ref{japos3} with the same landscape of 1st order phase transitions. 
The main difference of these two latter values appears in the finite jump of $\eta$, implying now a 1st order phase 
transition towards the parity breaking phase.

\begin{figure}[h!]
\centering
\includegraphics[scale=0.45]{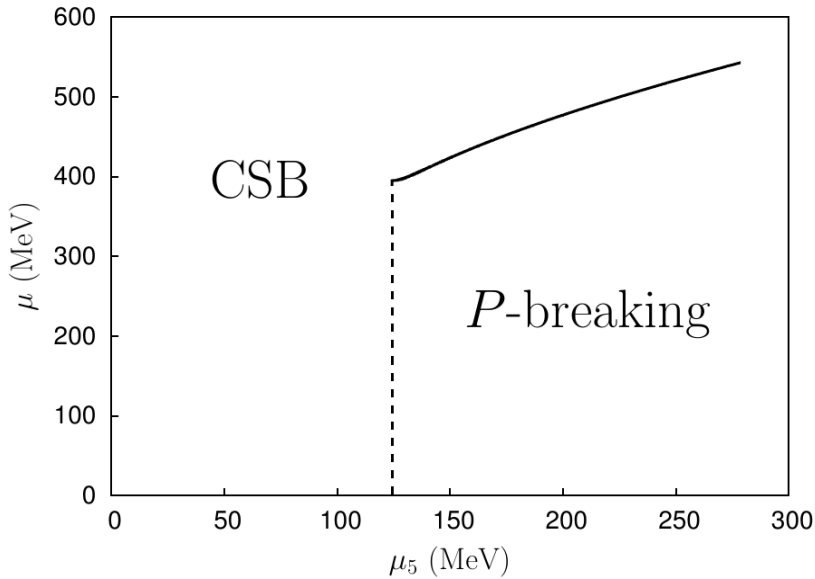}
\caption{Transition line from the CSB to the $P$-breaking phase with $G_1=-40/\Lambda^2$, $G_2=-39.5/\Lambda^2$, 
$m=-5$ MeV and $\Lambda=1$ GeV. The vertical dashed line is related to a 2nd order phase transition 
while the solid one corresponds to a 1st order one.}\label{mu5-mu}
\end{figure}

Finally, we present the phase transition line in a $\mu^c(\mu_5^c)$ plot in Fig. \ref{mu5-mu}. For $\mu<M^c\approx 395$ MeV 
(or equivalently, for $\mu_5^c=\mu_5^c(\mu=0)$), the transition is smooth (2nd order) while beyond that 
there is a jump in the condensates (1st order), as it was also observed in the previous figure.

\section{Conclusions and outlook}
The Nambu--Jona-Lasinio model has traditionally received much attention as a toy model for QCD. In spite of the
obvious shortcomings of this analogy, NJL is regarded as providing an intuitive picture  
of the mechanism of chiral symmetry breaking in QCD via a strong effective interaction in the scalar isosinglet 
channel. More recently the NJL model has received attention as a simpler arena where other aspects of QCD
could be tested, such as extreme QCD. Although it is far from obvious that NJL is a good modellisation of QCD then, 
these tests are still useful to understand in a simpler theory what are the right questions to pose.

In this context, the NJL model has been used recently by some authors \cite{azcoiti} to investigate the nature of the 
Aoki phase in QCD \cite{aoki}. This is a phase in lattice QCD with Wilson fermions where parity and possibly isospin 
symmetry is broken. It does not survive the continuum (note that the NJL does not have a `continuum limit' either).
It is however conceivable that the introduction of the chemical potential may enlarge the scope of the Aoki phase
and allow for a sensible continuum interpretation. This is what should happen if the effective theory analysis of
some of the present authors described in \cite{anesp} is correct. Finite chemical potential simulations being notoriously difficult in
lattice QCD, it is worth to analyse simpler theories such as NJL where analytical methods are available in the large $N$ limit.

The generation of an axial charge in heavy ion collision processes has also been contemplated in the theory. The effects
on QCD phenomenology of such a charge have been barely considered in the past. NJL may provide a first guidance 
to the problem too.

In this paper we work in the continuum and explore in detail the different phases that arise in the Nambu--Jona-Lasinio 
model in the presence of both vector and axial chemical potentials at zero temperature. The incorporation of $\mu_5$ together with $\mu$ had not been investigated before. The axial chemical potential changes considerably the thermodynamical 
properties of the model. It leads to a non-trivial dependence of the scalar condensate in the chirally broken phase. 
Interestingly, when the full $U(2)_L\times U(2)_R$ global symmetry is broken to $SU(2)_L\times SU(2)_R \times U(1)_V$ 
(i.e. $G_1 \neq G_2$) a phase where parity is spontaneously broken by the presence of an isosinglet condensate  
$\eta$ appears. However, we have not found any phase where parity and flavour symmetry are simultaneously broken thus
indicating the presence of a non-zero value for $\langle \bar\psi \gamma_5 \tau^3\psi\rangle$. On the contrary we have found
an extremely small region in the $\mu-\mu_5$ space of parameters where flavour symmetry is broken by a non-zero value of
$\langle \bar\psi\tau^3\psi\rangle$ but parity is not broken yet. However, the appearence of a parity breaking
condensate in the isosinglet sector is rather generic for $m\neq 0$.  

Demanding stability of such a phase however leads to a region of parameter space where the spectrum has little resemblance
to the one of QCD. We have investigated all the properties of the transition from the parity-even to the parity-odd phase 
providing results on the evolution of both condensates, which exhibit finite jumps under certain conditions, 
and finally examining the phase transition line, where it was shown that for $\mu<M^c$ we have a 2nd order transition 
while for $\mu>M^c$, it corresponds to a 1st order one.

The discussion presented here on the phase structure of the NJL model in the presence of external chemical potentials is
rather general and, as discussed above, the model -in spite of its simplicity- exhibits an enormously rich phase structure. 
This hopefully indicates that QCD still holds many surprises for us too.

\section*{Acknowledgements}
We would like to thank V. Azcoiti and E. Follana for numerous discussions concerning parity breaking in the NJL model 
and, particularly, for clarifying to us several points on the reality and positivity properties of the
fermion determinant. 
We acknowledge the financial support from projects FPA2010-20807, 2009SGR502, CPAN (Consolider CSD2007-00042).
A. A. Andrianov is also supported by Grant RFBR project 13-02-00127, Grant RFBR project 14-02-00095 as well
as by the Saint Petersburg State University grant 11.38.660.2013.
X. Planells acknowledges the support from Grant FPU AP2009-1855.

\appendix
\section{Calculation of the fermion determinant}\label{trick}
In this appendix we address the analysis of the determinant of the fermion operator presented in Eq. \eqref{fermoper}
\[\mathcal M(\mu,\mu_5)=\partial\!\!\!\!\!\!\not\;\;+(M+\vec\tau\vec a)-\mu\gamma_0-\mu_5\gamma_0\gamma_5+i\gamma_5(\vec\tau\vec\pi+\eta).\]
As it has been already stressed in \cite{azcoiti}, the fermion determinant can be proven to be real. The presence of both a vector and an axial chemical potentials does not modify this feature. Invariance under parity and time reversal symmetries also provide some equalities that will be useful for our purposes
\[\text{det}(\mathcal M(\mu,\mu_5))=\text{det}(\mathcal M^\dagger(\mu,\mu_5))=\text{det}(\mathcal M(\mu,-\mu_5))=\text{det}(\mathcal M^\dagger(\mu,-\mu_5)).\]
We shall use $N$ to be even in order the determinant to be positive defined and use the fact that $\det(\mathcal M)^2=\det(\mathcal M^2)$. The development of the product
\begin{align}
\nonumber &\mathcal M(\mu,\mu_5)\mathcal M^\dagger(\mu,-\mu_5)\\
\nonumber &=-\partial^2+M^2+\vec\pi^2+ (\eta^2+\vec a^2)+2 M\vec\tau\vec a+2 \eta\vec\tau\vec\pi+2 \gamma_5(\vec a\times \vec \pi)\vec\tau-\mu^2+\mu_5^2+2\mu\partial_0-2\mu_5\gamma_0\vec \gamma\vec \partial\gamma_5
\end{align}
provides a result which is scalar in flavour except for the term proportional to $\mu_5$. An additional product produces
\begin{align}
\nonumber \mathcal M(\mu,\mu_5)\mathcal M^\dagger(\mu,-\mu_5)\mathcal M(\mu,-\mu_5)\mathcal M^\dagger(\mu,\mu_5)=A'+\vec\tau(\vec \alpha'+\vec{\epsilon'}\gamma_5)
\end{align}
with
\begin{gather}
\nonumber A'=A^2+\vec\alpha^2+\vec\epsilon^2+4\mu_5^2\vec \partial^2, \qquad \vec \alpha'=2A\vec \alpha, \qquad \vec{\epsilon'}=2A\vec{\epsilon}\\
\nonumber A=-\partial^2+M^2+\vec\pi^2+ (\eta^2+\vec a^2)-\mu^2+\mu_5^2+2\mu\partial_0,\\
\nonumber \vec \alpha= 2 (M\vec a+\eta\vec\pi),\\
\nonumber \vec \epsilon= 2 (\vec a\times \vec \pi), \qquad \vec\alpha\vec{\epsilon}=0
\end{gather}
with the property $\vec \alpha'\vec{\epsilon'}=0$. The logarithm of a quantity with such characteristics can be calculated and all the non-diagonal operators in Dirac or flavour space disappear leading to
\begin{equation}
\nonumber \log[A+\vec\tau(\vec \alpha+\vec{\epsilon}\gamma_5)]=\frac12\log[A^2-\vec \alpha^2-\vec{\epsilon}^2].
\end{equation}
The evaluation of the argument leads us to
\begin{align}
\nonumber &A'^2-\vec \alpha'^2-\vec{\epsilon'}^2=\prod_{\pm} \left [-(ik_0+\mu)^2+(|\vec k|\pm \mu_5)^2+M_+^2\right ]\left [-(ik_0+\mu)^2+(|\vec k|\pm \mu_5)^2+M_-^2\right ]
\end{align}
where $M_\pm^2=(M\pm a)^2+(\eta\pm\pi)^2$. Finally the fermion determinant can be written as
\begin{align}
\log\text{det}(\mathcal M(\mu,\mu_5))=&\text{Tr}\log\mathcal M(\mu,\mu_5)=\frac18\text{Tr}\log(A'^2-\vec \alpha'^2-\vec{\epsilon'}^2)\\
\nonumber =&\frac18\text{Tr}\sum_{\pm}\Bigg\{\log\left [-(ik_0+\mu)^2+(|\vec k|\pm \mu_5)^2+M_+^2\right ]\\
\nonumber &+\log\left [-(ik_0+\mu)^2+(|\vec k|\pm \mu_5)^2+M_-^2\right ]\Bigg\},
\end{align}
where the trace operator is given by
\[\text{Tr}(1)=8NT\sum_n\int\frac{d^3\vec k}{(2\pi)^3}[k_0\to \omega^F_n],\]
with $\omega^F_n=(2n+1)\pi/\beta$.

\end{document}